# Lorentz-boost eigenmodes

Konstantin Y. Bliokh[1,2]

[1]*Nonlinear Physics Centre, RSPE, The Australian National University, Canberra, Australia*
[2]*Center for Emergent Matter Science, RIKEN, Wako-shi, Saitama 351-0198, Japan*

Plane waves and cylindrical or spherical vortex modes are important sets of solutions of quantum and classical wave equations. These are eigenmodes of the energy-momentum and angular-momentum operators, i.e., generators of *spacetime translations* and *spatial rotations*, respectively. Here we describe another set of wave modes: eigenmodes of the "boost momentum" operator, i.e., a generator of *Lorentz boosts* (spatio-temporal rotations). Akin to the angular momentum, only one (say, $z$) component of the boost momentum can have a well-defined quantum number. The boost eigenmodes exhibit invariance with respect to the Lorentz transformations along the $z$-axis, leading to scale-invariant wave forms and step-like singularities moving with the speed of light. We describe basic properties of the Lorentz-boost eigenmodes and argue that these can serve as a convenient basis for problems involving causal propagation of signals.

## 1. Introduction

Symmetries are one of the cornerstones of modern physics [1]. The Poincaré group of symmetries of the Minkowski spacetime plays a special role, because the most fundamental conservation laws of energy, momentum, and angular momentum are associated with these symmetries via the celebrated Noether's theorem [1–3]. The Poincaré group is 10-dimensional, containing: (i) *space-time translations* (4 dimensions), (ii) *spatial rotations* (3 dimensions) and (iii) spatio-temporal rotations, i.e., *Lorentz boosts* (3 dimensions) [4,5]. The *energy-momentum* and *angular-momentum* conservation laws, associated with the symmetries (i) and (ii) are well known and widely used in physics. The remaining conserved quantities produced by the Lorentz-boosts symmetry (iii) form a vector, which can be called "*boost momentum*" [6–10], and is much less in demand in physical problems.

In quantum mechanics or wave physics (we consider the first-quantization wave approach), physical quantities become *operators*, and now these operators and their *eigenmodes* play a fundamental role. In particular, the operators of the 10 conserved quantities mentioned above are generators of the corresponding Poincaré transformations of the spacetime [5,7,9,11,12]. Moreover, eigenmodes of these operators form the most useful sets of wave eigenmodes, which can be associated with the corresponding spacetime symmetries. Namely, eigenmodes of the energy-momentum operator (i) are *plane waves*, while eigenmodes of the angular momentum operator (ii) (in fact, its projection on a chosen axis or squared angular momentum) are cylindrical or spherical *vortex modes* [13–15].

From this consideration, a natural question arises: what are the wave eigenmodes of the remaining boost-momentum operator (iii)? In this paper, we examine this set of *Lorentz-boost eigenmodes*, which also satisfy the wave equation (we consider the scalar Klein-Gordon equation). We describe basic properties of these modes, which are intimately related to the Lorentz-boost invariance, and discuss physical problems where such modes can appear.

While plane waves are delocalized modes without singularities, the vortex modes are better localized (still being unbounded in space) and possess *phase singularities* [16–18]. A distinctive feature of the Lorentz-boost eigenmodes is that they can *vanish identically* in the $z>t$ or $z<-t$ regions of spacetime, and, as a consequence, possess essential *phase and amplitude singularities*.

Such modes are well suited for problems involving *causal signal propagation* [19–21]. Indeed, even localized (i.e., square-integrable but spatially unbounded) wave packets made of plane waves bring about paradoxes with the superluminal propagation [22–24]. In contrast, boost eigenmodes immediately provide step-like singularities (signals) propagating *exactly with the speed of light* and never violating causality. Remarkably, this speed-of-light propagation of the boost eigenmodes is independent of mass (even for the imaginary-mass tachyons), which is in agreement with the luminal propagation of the Sommerfeld signal precursors even for massive relativistic particles or dispersive optical media [19–21].

Similar, but not the same, set of Lorentz-boost eigenmodes have been considered in the context of "point-form" quantum field theory [25–28]. The difference is that the previous solutions [25–28] were obtained to satisfy 1D Klein-Gordon equation, and these are initially defined only inside the forward light cone; extending these solutions to the whole Minkowski spacetime faces some difficulties. In contrast, the modes considered in this work are solutions of the 1D *massless* wave equation (the mass term is taken into account in the equation for the transverse wavefunction envelope), and our modes are immediately well-defined in the *whole* Minkowski spacetime.

## 2. Poincaré-group symmetries, operators, eigenmodes

Throughout this work we use units $c = \hbar = 1$ and consider scalar waves $\psi(t,\mathbf{r})$ in the Minkowski spacetime, which satisfy the Klein-Gordon relativistic wave equation [29]:

$$\left[\partial_t^2 - \nabla^2 + \mu^2\right]\psi(t,\mathbf{r}) = 0. \qquad (1)$$

Here $\mu$ is the mass, and our consideration below includes massive, $\mu^2 > 0$, massless, $\mu = 0$, and tachyon, $\mu^2 < 0$ cases. We will describe the wave function via its "intensity" $I = |\psi|^2$, phase $\mathrm{Arg}\,\psi$, and real part $\mathrm{Re}\,\psi$ involving both the amplitude and phase, whereas the conserved density of particles (minus the density of antiparticles, so it is not positive-definite) and the corresponding current for the Klein-Gordon equation are given by [29]:

$$\rho = -\mathrm{Im}\left(\psi^* \partial_t \psi\right) \neq I, \quad \mathbf{j} = \mathrm{Im}\left(\psi^* \nabla \psi\right) = I \nabla \mathrm{Arg}\,\psi. \qquad (2)$$

There are 10 Poincaré symmetries, including spacetime translations, spatial rotations, and Lorentz boosts. The corresponding generators are operators of the energy-momentum, $(\hat{E}, \hat{\mathbf{p}})$, angular momentum, $\hat{\mathbf{L}}$, and the boost momentum, $\hat{\mathbf{N}}$ [3,5,9–12], which satisfy the following commutation relations [4,5,11]:

$$[\hat{E},\hat{P}_i] = [\hat{E},\hat{L}_i] = [\hat{P}_i,\hat{P}_j] = 0, \quad [\hat{L}_i,\hat{P}_j] = i\varepsilon_{ijk}\hat{P}_k, \quad [\hat{E},\hat{N}_i] = i\hat{P}_i, \quad [\hat{P}_i,\hat{N}_j] = i\delta_{ij}\hat{E},$$

$$[\hat{L}_i,\hat{L}_j] = -[\hat{N}_i,\hat{N}_j] = i\varepsilon_{ijk}\hat{L}_k, \quad [\hat{L}_i,\hat{N}_j] = i\varepsilon_{ijk}\hat{N}_k, \qquad (3)$$

where $\varepsilon_{ijk}$ and $\delta_{ij}$ are the Levi-Civita symbol and Kronecker delta, respectively.

First, the energy-momentum operators are:

$$\left(\hat{E},\hat{\mathbf{p}}\right) = \left(i\partial_t, -i\nabla\right). \qquad (4)$$

Their eigenmodes are *plane waves*:

$$\psi_{\mathbf{k}} \propto \exp\left[i(\mathbf{k}\cdot\mathbf{r} - \omega t)\right]. \qquad (5)$$

Here, the frequency and wave vector $(\omega, \mathbf{k}) \in \mathbb{R}^4$ are the corresponding eigenvalues, and the dispersion relation $\omega^2 - k^2 - \mu^2 = 0$ holds to satisfy the wave equation (1). Note that, for plane waves (5), the intensity $I$ and particle density $\rho$ are invariant with respect to spacetime translations, while the wavefunction $\psi$ only acquires a phase shift:

$$(t, \mathbf{r}) \to (t + \delta t, \mathbf{r} + \delta \mathbf{r}), \quad \psi_{\mathbf{k}} \to \psi_{\mathbf{k}} \exp\left[i(\mathbf{k} \cdot \delta \mathbf{r} - \omega \delta t)\right]. \tag{6}$$

Note also that, assuming propagation along the $z$-axis, $\mathbf{k} \cdot \mathbf{r} = k_z z$, the phase velocity of plane waves (5) is *superluminal*: $|v_{\mathrm{ph}}| = |\omega / k_z| > 1$ for massive waves with $\mu > 0$. This does not contradict causality because a pure plane wave does not transport any signal [21]. The $z$-propagating plane-wave wavefunctions are shown in Fig. 1a.

Second, the angular-momentum operator is

$$\hat{\mathbf{L}} = \mathbf{r} \times \hat{\mathbf{p}} = -i \mathbf{r} \times \nabla. \tag{7}$$

Since its components do not commute with each other, as seen from Eqs. (3), one can choose eigenmodes of only one component, say, $\hat{L}_z = i(y \partial_x - x \partial_y) = -i \partial_\varphi$, where $\varphi$ is the azimuthal angle of the cylindrical or spherical coordinates attached to the $z$-axis. The eigenmodes of $\hat{L}_z$ are *vortex modes* [13–15]:

$$\psi_m \propto [x + i \operatorname{sgn}(m) y]^{|m|} \propto \exp(im\varphi), \tag{8}$$

where $m \in \mathbb{Z}$ is the integer eigenvalue. The dependence of the vortex modes (8) on other coordinates and time can be different. Since, according to Eqs. (3), $\hat{L}_z$ commutes with $\hat{E}$, $\hat{P}_z$, and $\hat{L}^2$ (but $\hat{P}_z$ does not commute with $\hat{L}^2$), usually one chooses the vortex modes to also be the eigenmodes of $\hat{E}$ and $\hat{P}_z$ (monochromatic cylindrical beams) [14,15,30] or $\hat{E}$ and $\hat{L}^2$ (monochromatic spherical modes) [13]. For example, the cylindrical vortex modes are

$$\psi_{m \omega k_z} \propto R_m(r_\perp) \exp\left[i(m\varphi + k_z z - \omega t)\right], \tag{9}$$

where $(r_\perp, \varphi, z)$ are the cylindrical coordinates, whereas $R_m(r_\perp) = J_{|m|}(k_\perp r_\perp)$ and $\omega^2 - k_z^2 - k_\perp^2 - \mu^2 = 0$ to satisfy the wave equation (1) [8]. Akin to the plane-wave invariance with respect to translations, the vortex modes (8) and (9) have intensities and probability densities invariant with respect to the rotations about the $z$-axis. The wavefunctions acquire phase factors upon such rotations:

$$\varphi \to \varphi + \delta\varphi, \quad \psi_m \to \psi_m \exp(im\delta\varphi). \tag{10}$$

An important peculiarity of the vortex modes is that they contain *phase singularities* (indeterminate phases and divergent phase gradients) at the vanishing-intensity point $x + iy = 0$, i.e., $r_\perp = x = y = 0$ [16–18]. These singularities are points in the transverse 2D plane orthogonal to the angular-momentum direction. Examples of vortex wavefunctions (8) are shown in Fig. 1b.

Finally, the remaining operator of the boost momentum (consisting of the three generators of Lorentz boosts) reads [6–12]:

$$\hat{\mathbf{N}} = t\hat{\mathbf{p}} - \mathbf{r}\hat{E} = -i(t\nabla + \mathbf{r}\partial_t). \tag{11}$$

Similarly to the angular momentum, noncommuting components of $\hat{\mathbf{N}}$ allow us to consider the eigenmodes of only one component, $\hat{N}_z = -i(t\partial_z + z\partial_t)$. Note that $\hat{N}_i$ commutes with the d'Alembertian operator $\left(\hat{P}^2 - \hat{E}^2\right)$ of the wave equation (1), and these modes can be solutions of the Klein-Gordon equation. It is not difficult to see that the eigenmodes of $\hat{N}_z$ are:

$$\psi_n \propto \left[-t + \text{sgn}(n)z\right]^{-i|n|} = \exp\left[-i|n|\ln\left(-t + \text{sgn}(n)z\right)\right], \qquad (12)$$

where $n \in \mathbb{R}$ is the corresponding boost-momentum eigenvalue (with a continuous spectrum). The *boost eigenmodes* (12) are the central object of the present study. Examples of the wavefunctions (12) are shown in Fig. 1c.

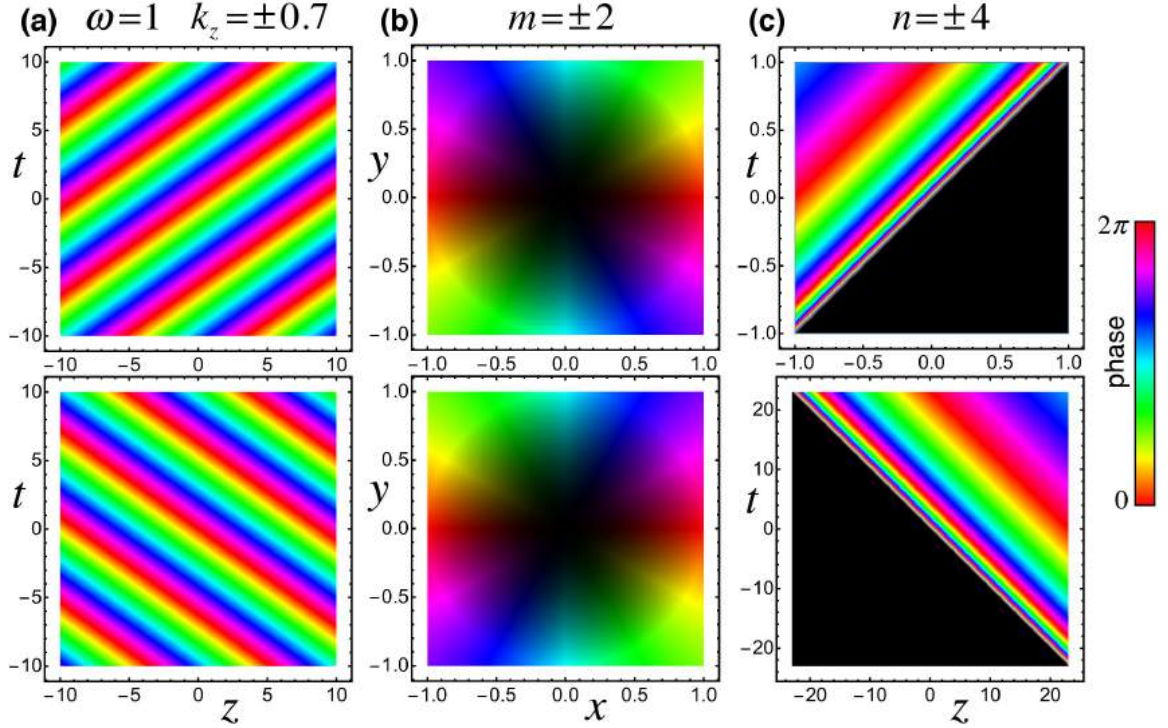

**Fig. 1.** Plane waves **(a)**, vortex modes **(b)**, and boost eigenmodes **(c)**, Eqs. (5), (8), and (12), with positive (upper panels) and negative (lower panels) eigenvalues. Here, the mass is $\mu = 0.7$, and propagation along the $z$-axis is assumed for plane waves. Colors indicate the wavefunction phase $\text{Arg}\,\psi$, whereas brightness corresponds its absolute value $|\psi|$ [31]. In contrast to plane waves, the phase fronts of the boost modes, always correspond to phase velocity equal to the speed of light, independently of the mass $\mu$. The two panels **(c)** also demonstrate the *scale invariance* of the boost modes (notice the scales), which is a counterpart of the translation and rotation invariance of the plane waves and vortex modes.

One can see certain similarity with the vortex modes (8), but the boost eigenmodes reflect the *hyperbolic* geometry of the spacetime. Indeed, introducing the hyperbolic coordinates

$$z = u\sinh\phi, \qquad t = u\cosh\phi, \qquad (13)$$

the boost-momentum operator acquires becomes $\hat{N}_z = -i\partial_\phi$ (cf. the angular-momentum $\hat{L}_z = -i\partial_\varphi$), whereas the modes (12) exhibit a vortex-like dependence on the hyperbolic angle: $\psi_n \propto \exp(in\phi)$. It follows from here that the boost eigenmodes can be multiplied by any function of $u = \sqrt{t^2 - z^2}$:

$$\psi_n \to f\left(\sqrt{t^2 - z^2}\right)\psi_n. \tag{14}$$

For example, choosing $f = u^{i|n|}$ makes the functions (12) $u$-independent. Hyperbolic coordinates (13) and specific choice of the function $f(u)$ was used in the point-form quantum field-theory approach [25–28]. However, this hyperbolic formalism has a drawback: real values of $(u,\phi)$ cover only $t^2 - z^2 > 0$ part of the spacetime (inside the light cone). Moreover, only the forward light cone mapped to the $u > 0$ values is usually considered. Thus, the extension of the Lorentz-boost eigenmodes considered in [25–28] to the whole spacetime requires some care: "altogether it does not seem to be very practical to study evolution of quantum field theories in hyperbolic coordinates" [28]. In contrast, the modes (12) are well defined on the whole Minkowski spacetime $(z,t)$.

Being functions of $(z+t)$ or $(z-t)$, the boost eigenmodes (12) satisfy the 1D *massless* wave equation: $(\partial_t^2 - \partial_z^2)\psi_n(z,t) = 0$. There are two ways to satisfy the Klein-Gordon equation (1). First, the $(z,t)$-dependent modes (12) can be multiplied by a solution of the 2D Helmholtz-like equation $(\nabla_\perp^2 - \mu^2)\chi(x,y) = 0$, $\nabla_\perp^2 = \partial_x^2 + \partial_y^2$. In this approach, the mass term is taken into account by the transverse wavefunction $\chi(x,y)$. Second, one can modify the boost eigenmodes to satisfy the 1D *massive* Klein-Gordon equation $(\partial_t^2 - \partial_z^2 + \mu^2)\psi_n(z,t) = 0$ using the substitution (14) with suitable function $f(t^2 - z^2)$. This approach was realized in [25–28]. The ambiguity originates from the fact that, separating the variables, one can ascribe the mass term either to the transverse $(x,y)$ or to longitudinal $(z,t)$ coordinates, or to split it between these degrees of freedom. These solutions of the full Klein-Gordon equation (1) are considered in Section 3.6 below.

Most importantly, the boost eigenmodes (12) have nonuniform distributions of the intensity $I_n = |\psi_n|^2$ with *step-like singularities*, which propagate with the speed of light in the forward ($n > 0$) or backward ($n < 0$) $z$-direction. In fact, the functions (12) have multivalued amplitudes, which can be fixed by choosing one branch of the complex logarithm function. Choosing the main branch $\ln a = \ln|a| + i \operatorname{Arg} a$ and introducing normalizing amplitudes $\psi_n \to \exp(-\pi|n|)\psi_n$, we have

$$I_n \propto \begin{cases} 1, & z < t \\ e^{-2\pi|n|}, & z > t \end{cases} \text{ for } n > 0 \quad \text{and} \quad I_n \propto \begin{cases} e^{-2\pi|n|}, & z < -t \\ 1, & z > -t \end{cases} \text{ for } n < 0. \tag{15}$$

Thus, one can associate the eigenmodes (12) with the *causal propagation of step-like signals* [21]. Moreover, in contrast to plane waves, the local phase velocity in the boost eigenmodes always equals to the speed of light. For the signal-propagation problems, it makes sense to set the exponentially-small wavefunction amplitudes to zero in the zones $z > t$ ($n > 0$) and $z < -t$ ($n < 0$). This cut-off procedure is realized via multiplication by the Heaviside step function:

$$\psi_n \to \psi_n \Theta[t - \operatorname{sgn}(n)z]. \tag{16}$$

It is easy to verify that the wavefunctions (16) are still the eigenmodes of $\hat{N}_z$.

In contrast to non-singular plane waves and vortices with point phase singularities, both the amplitude and phase are singular in the boost eigenmodes for $\operatorname{sgn}(n)z = t$, and these singularities are lines in the $(z,t)$ plane. Such highly singular behavior is essential for causal signal-propagation problems, because vanishing of the wavefunction in the half-plane $\text{sgn}(n)z - t > 0$ requires it to be non-analytical for $\text{sgn}(n)z = t$.

Obviously, the step-like intensity distributions moving with the speed of light are invariant with respect to the Lorentz boosts along the $z$-axis. Furthermore, such Lorentz transformations produce the phase shift in the boost eigenmodes (12), which is quite similar to Eqs. (6) and (10) for plane and vortex waves. Indeed, the Lorentz boost is a hyperbolic rotation described by

$$(z' - t') = (z - t)e^{\delta\phi}, \quad (z' + t') = (z + t)e^{-\delta\phi}, \tag{17}$$

where $\delta\phi = \tanh^{-1} v_z$ is the rapidity corresponding to the motion of the reference frame with velocity $v_z$. Thus, the Lorentz boost shifts the hyperbolic angle (13), $\phi = \phi' + \delta\phi$, resulting in the corresponding phase shift in the boost modes (12):

$$\phi \to \phi + \delta\phi, \quad \psi_n \to \psi_n \exp(in\delta\phi). \tag{18}$$

Notably, since the boost eigenmodes (12) are functions of $(z+t)$ or $(z-t)$, the invariance with respect to the Lorentz boosts (17) means the *scale invariance* of the wavefunctions $\psi_n$. This is clearly seen in the two panels of Fig. 1c plotted on different scales. Below we describe properties of the boost eigenmodes (12) in more details.

## 3. Properties of the Lorentz-boost eigenmodes

*3.1. Orthogonality*

It is known that the plane waves and vortices form complete sets of mutually orthogonal wave modes: $\langle \psi_{\mathbf{k}} | \psi_{\mathbf{k}'} \rangle \propto \delta(\mathbf{k} - \mathbf{k}')$ and $\langle \psi_m | \psi_{m'} \rangle \propto \delta_{mm'}$. It might seem that the boost eigenmodes are not mutually orthogonal because the integral $\int \psi_n^* \psi_{n'} dz$ does not yield the delta-function $\delta(n - n')$. There are two solutions to this problem.

First, if we introduce the inner-like product determined by the form of the particle density (2), $\langle \psi | \psi' \rangle \equiv -\int \text{Im}(\psi^* \partial_t \psi') dz$, the boost modes (12) become mutually orthogonal:

$$\langle \psi_n | \psi_{n'} \rangle \propto \delta(n - n'). \tag{19}$$

Indeed, for the pairs of modes with the same $\text{sgn}(n)$, the inner product is determined (at $t = 0$) by the integral $\int z^{-1} \exp[i(n - n')\ln z] dz \propto \delta(n - n')$, while for modes with the opposite $\text{sgn}(n)$, their intensities (16) have zero overlap at $t = 0$. Note that changing the inner product from $\int \psi^* \psi' dz$ to $-\int \text{Im}(\psi^* \partial_t \psi') dz$ does not affect the inner products of monochromatic plane waves or vortex modes; for monochromatic modes these products differ by the constant factor $\omega'$. However, the modified product $\langle \psi | \psi \rangle$ is not positive-definite because $\rho$ is not positive-definite. This is related to the spin-0 nature of the Klein-Gordon particles; for spin-1/2 Dirac particles, both $\rho = |\psi|^2$ and the corresponding $\langle \psi | \psi \rangle = \int |\psi|^2 dz$ become positive-definite.

Second, since the boost momentum and its eigenmodes are related to the spacetime rotations, it is natural to introduce the inner product using the spacetime integration: $\langle \psi | \psi' \rangle \equiv \int \psi^* \psi' dz dt$. Here $\langle \psi | \psi \rangle$ is positive-definite and the orthogonality (19) immediately follows from the hyperbolic-coordinate (13) representation: $\psi_n(u, \phi) \propto \exp(in\phi)$, $dz dt = -u\, du\, d\phi$.

## 3.2. Particle density, current, and Fourier spectrum

We now consider the wavefunctions (12) in more detail. Figure 2a shows distributions $\text{Re}\,\psi_n(0,z)$ and $|\psi_n(0,z)|$ for modes with different eigenvalues $n>0$ ($n<0$ modes have the corresponding mirror-symmetric profiles). One can see that the number of half-oscillations visible in the plots approximately equals $|n|$. The local wavevector diverges hyperbolically near the wavefunction singularity $z=\text{sgn}(n)t$. This makes the visible wavefunction profile scale-invariant: zooming-in the wavefunction reveals more oscillations near the singularity. Note that the increasing local frequency of oscillations near the signal front is a typical feature of the *Sommerfeld signal precursors*, always propagating with the speed of light [19–21].

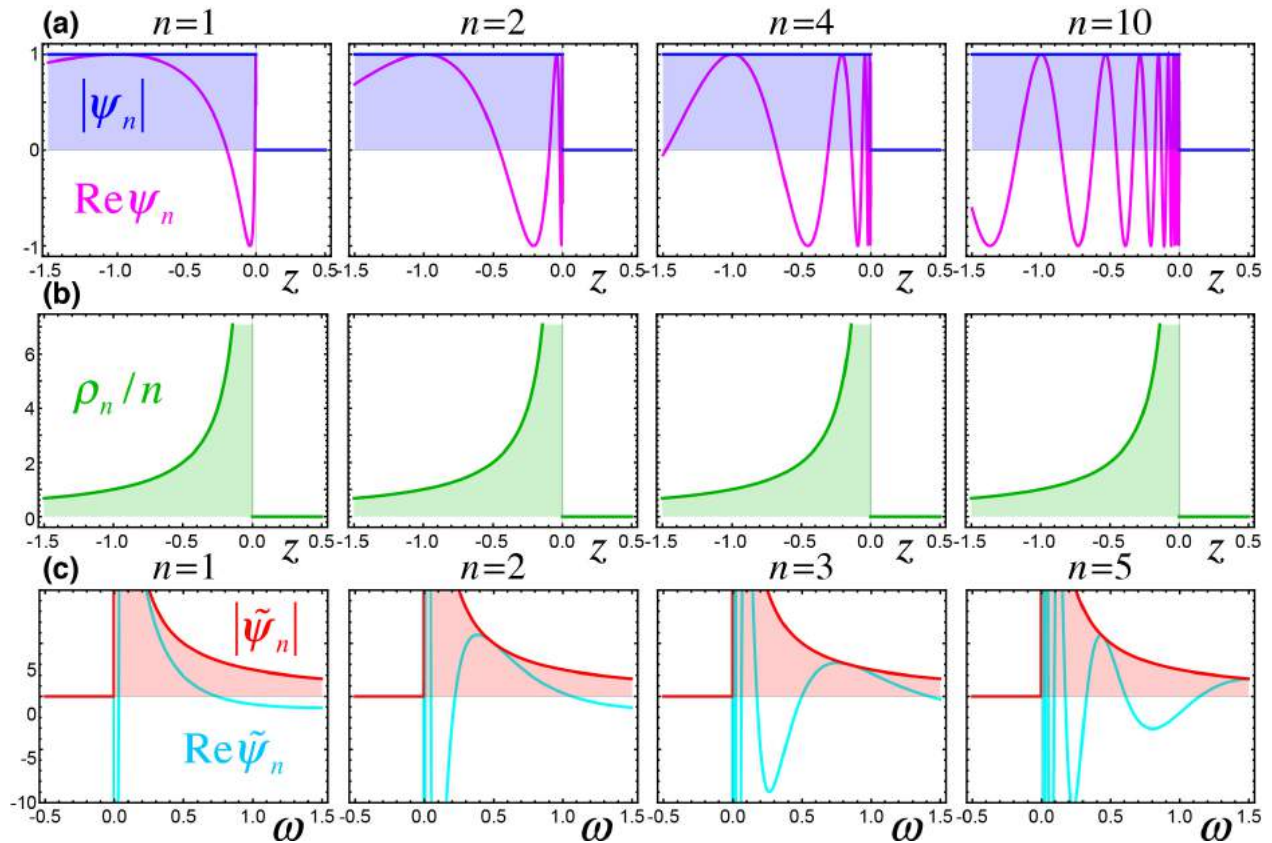

**Fig. 2.** (a) The boost eigenmodes (12) $\psi_n(0,z)$ with different boost-momentum eigenvalues $n>0$. The number of oscillations visible in the plots approximately corresponds to $n$, independently of the $z$-scaling. (b) The particle density and current (20) for the boost eigenmodes (12). (c) Fourier spectra of the boost eigenmodes with different eigenvalues $n$ (here the normalized amplitudes $\tilde{\psi}_n(\omega)/\sqrt{n}$ are plotted). The phase and amplitude behavior of the Fourier components (with singularity at $\omega=0$) is similar to the behavior of the wavefunctions **(a)** and probability densities **(b)**, respectively.

The particle density and current (2) in the boost eigenmodes (12) are characterized by the local temporal and spatial gradients of the wavefunction phase. These are given by

$$\rho_n \propto \frac{|n|\,I_n}{t-\text{sgn}(n)z}, \qquad j_{zn} = \text{sgn}(n)\rho_n. \tag{20}$$

Thus, in contrast to the step-like distribution of the "intensity" $I_n$, the particle density and current have a hyperbolic dependence diverging near the wavefunction singularity and decaying away from it, see Fig. 2b. Note that, considering the particle current and density as "local wavevector" and "local frequency" [32], the *local phase velocity* is given by their ratio, and its absolute value always equals to the speed of light. This is in contrast to the superluminal phase velocity of plane waves (5) [21].

The presence of the singularity and scale-invariant behavior implies unbounded Fourier spectrum of the boost eigenmodes. The spatial and temporal frequencies coincide (up to the sgn($n$) factor), and the Fourier components of the function $\psi_n(-t+\text{sgn}(n)z) \equiv \psi_n(\zeta)$ can be calculated as

$$\tilde{\psi}_n(\omega) = \int \psi_n(\zeta) e^{-i\omega\zeta} d\zeta \propto \int e^{-i|n|\ln\zeta - i\omega\zeta} d\zeta . \tag{21}$$

Fourier spectra (19) for the boost eigenmodes with different $n$ are plotted in Fig. 2c. One can see that the Fourier spectra have hyperbolic-like envelopes diverging for $\omega = 0$ and decaying for $\omega \to \infty$. Moreover, due to the singular character of the boost eigenmodes, their Fourier spectra unavoidably have *nonzero negative-frequency (negative-energy) amplitudes*: $\tilde{\psi}_n(\omega < 0) \neq 0$. These are exponentially small and not visible in Fig. 2, but crucial for the causal propagation [21,33,34]. The phase structure of the Fourier spectra (21) resembles the scale-invariant behavior of the wavefunctions $\psi_n$, with the density of oscillations diverging near the $\omega = 0$ singularity.

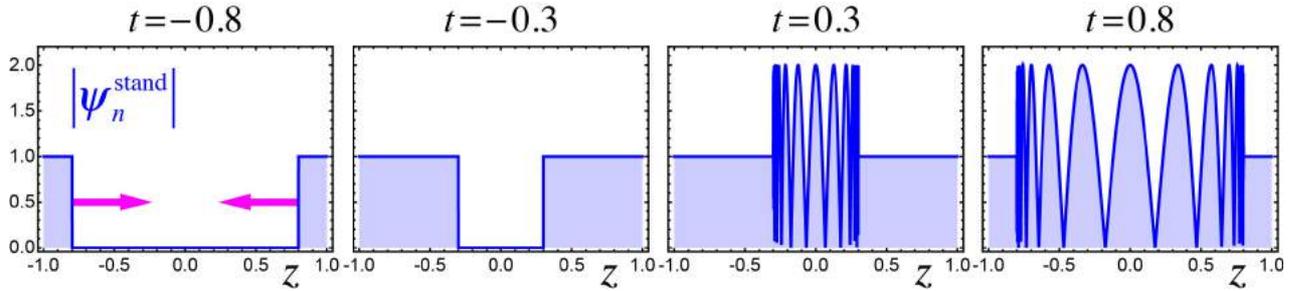

**Fig. 3.** Temporal evolution of the spatial intensity distribution of the standing mode (22) and (23) with $n = 7$.

### 3.3. Standing waves

Superpositions of plane waves with opposite wavevectors $\pm \mathbf{k}$, or vortices with opposite angular-momentum quantum numbers $\pm m$, form standing waves. In a similar manner, one can consider the standing-wave superpositions of the boost eigenmodes (12) with opposite eigenvalues $\pm n$:

$$\psi_n^{\text{stand}} \propto (-t+z)^{-i|n|} + (-t-z)^{-i|n|} . \tag{22}$$

These solutions describe two counter-propagating step-like signals, which meet at $t = 0$ and then start to interfere, as shown in Fig. 3. The number of visible maxima in the interference region approximately corresponds to $|n|$. The intensity of the standing wave (22) can be expressed via intensities of the propagating modes (12) as

$$I_n^{\text{stand}} = I_n + I_{-n} + 2\sqrt{I_n I_{-n}} \cos\left[n\left(\ln|z-t| - \ln|z+t|\right)\right]. \tag{23}$$

Interestingly, the temporal dynamics of the spatial distribution $|\psi_n^{\text{stand}}(z)|$ in the standing modes (22) looks like horizontal scaling (stretching or squeezing) of the function. Note also that the intensity distributions in the standing plane waves and standing vortex waves are invariant with respect to *discrete* translations (6) and rotations (10), respectively: $\delta z = \pi/k$ and $\delta\varphi = \pi/m$.

Similarly, the intensity of the standing boost mode (22) and (23) is invariant with respect to discrete Lorentz transformations (17) and (18) with $\delta\phi = \pi/n$.

*3.4. Wave packets*

By analogy with wave packets consisting of multiple plane waves with different frequencies/wavevectors, one can construct wave packets of the boost modes (12) with different eigenvalues $n$. The simplest option is to consider a superposition of modes with the Gaussian distribution, characterized by the central value $n_0$ and the width $\Delta n$:

$$\psi(-t + \mathrm{sgn}(n_0)z) = \int_0^{\mathrm{sgn}(n_0)\infty} \psi_n(-t + \mathrm{sgn}(n)z) \exp\left[-\frac{(n-n_0)^2}{\Delta n^2}\right] dn. \quad (24)$$

Here, assuming $\Delta n \ll |n_0|$, we chose the limits of integration corresponding to the modes with the same propagation direction, i.e., $\mathrm{sgn}(n)$. Obviously, the resulting wave packet is a function of $(-t + \mathrm{sgn}(n_0)z)$, i.e., propagates with the speed of light without distortions.

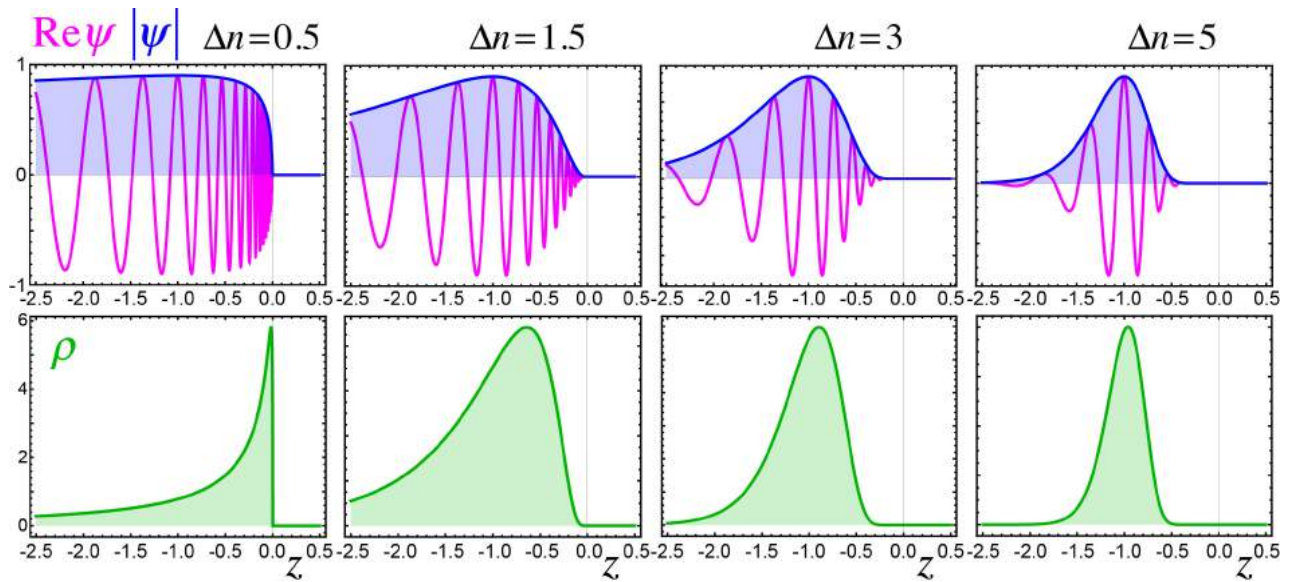

**Fig. 4.** Wavefunctions $\psi(0,z)$ and the corresponding probability densities $\rho(0,z)$ for the wave packets (24) of the boost eigenmodes with the central boost momentum $n_0 = 20$ and different values of the spread $\Delta n$.

Figure 4 shows examples of the wave packets (24) with $n_0 = 20$ and different values of $\Delta n$. Remarkably, these Gaussian-like boost wave packets *are free of step-like intensity discontinuities and diverging probability densities* present in pure boost eigenmodes (Fig. 2). Moreover, for $\Delta n \gg 1$, the wave packet (24) becomes well localized and looks like a usual Gaussian wave packet with the well-pronounced central frequency (wave vector). Thus, tuning $\Delta n$, one can observe transition from step-like scale-invariant signals (with all frequencies equally seen in the wavefunction oscillations) to Gaussian-like wave packets with the central frequency $\omega_c = |k_{zc}| \sim |n_0|$. However, the resulting Gaussian-like wavefunction still vanishes identically (assuming the cut-off (16)) for $\mathrm{sgn}(n_0)z - t > 0$, which is suitable for the causal-propagation problems [19–23]. The asymptotic form of the wavepacket (24) for $|\ln(-t + \mathrm{sgn}(n_0)z)| \gg 1$ can be obtained using the stationary-phase method:

$$\psi(\zeta) \propto \exp\left[-i|n_0|\ln|\zeta| - \frac{\Delta n^2}{4}\ln^2|\zeta|\right]\Theta(-\zeta), \quad \zeta = -t + \text{sgn}(n_0)z, \qquad (25)$$

where we used the cut-off (16). This asymptotic form perfectly reproduces the wave packets shown in Fig. 4. From Eq. (25), one can see that $\psi(0) = \psi'(0) = \psi''(0) = ... = 0$, i.e., the wavefunction is *smooth* (infinitely-differentiable) but still *non-analytical* in $\zeta = 0$.

Importantly, signals perfectly propagating with speed of light are typical only for 1D massless wave equations. In 3D wave problems, particularly with a finite real mass, the main parts of signals (excluding luminal precursors) propagate with subluminal speeds and experience distortions of their shapes [21]. Such solutions can be constructed either by interfering boost eigenmodes (12) with different $\text{sgn}(n)$, or by using the boost eigenmodes modified by a prefactor (14) (e.g., the one making the modes to satisfy the 1D Klein-Gordon equation [25–29]). This will result in a wave packet, which is not a function of $(-t + \text{sgn}(n_0)z)$; its shape will vary during propagation.

*3.5. Uncertainty relations*

Quantum-mechanical consideration of the generators of the Poincaré group reveals an interesting connection with the *uncertainty relations*. Namely, the operators of the energy-momentum are associated with the translational invariance of the spacetime, and the corresponding uncertainty relations between the energy-momentum values and time-coordinates appear: $\Delta\omega\,\Delta t \geq 1/2$ and $\Delta k_i\,\Delta r_i \geq 1/2$. For pure energy-momentum eigenmodes, i.e., plane waves, $\Delta\omega = \Delta k_i = 0$, and the time and position are indeterminate: $\Delta t = \Delta r_i = \infty$. One can interpret this as indistinguishability of two plane waves related by a spacetime translation. In a similar manner, there is an uncertainty relation between the angular-momentum quantum number $m$ and azimuthal coordinate: $\Delta m \Delta\varphi \geq 1/2$ [35,36]. For pure vortex modes $\Delta m = 0$, and the angular position $\varphi$ is indeterminate, i.e., vortex modes related by a $\varphi$-rotation are indistinguishable. (Note, however, that the angular uncertainty cannot exceed $\Delta\varphi_{\max} = \pi/\sqrt{3}$ due to the cyclic nature and limited range of values $(0, 2\pi)$, and this modifies the angular uncertainty relation [35,36].) By analogy, one can expect that there is a uncertainty relation for the boost eigenvalues:

$$\Delta n\,\Delta\phi \geq 1/2. \qquad (26)$$

Here, we recall that the hyperbolic angle $\phi$, Eq. (13), determines the *references frame* and the *velocity* of its motion along the $z$-axis. For pure boost eigenmodes, $\Delta n = 0$, and the observer cannot distinguish between the signals received in two reference frames moving with respect to each other: the wavefunction has the same form in any frame, and $\Delta\phi = \infty$. In contrast, for wave packets with $\Delta n \gg 1$, the wavefunction has a well-pronounced central frequency (Fig. 4), which will change depending on the reference frame. In this case, the observer can distinguish between the signals seen in the two reference frames characterized by relative rapidity $\Delta\phi \sim 1/\Delta n \ll 1$.

The uncertainty relations are described by commutation relations between the corresponding operators: $[r_i, \hat{P}_j] = i\delta_{ij}$, $[t, \hat{E}] = -i$, etc. For the boost momentum operator, one can expect similar non-commutativity with the *velocity operator*: $[\hat{v}_i, \hat{N}_j] \sim i\delta_{ij}$. Taking into account the velocity of a classical relativistic particle, $\mathbf{v} = \mathbf{P}/E$, and the commutation relation (3) between the momentum and boost momentum: $[\hat{P}_i, \hat{N}_j] = i\delta_{ij}\hat{E}$, we see that this is a reasonable hypothesis. Using the relativistic velocity operator $\hat{\mathbf{v}} = \hat{\mathbf{P}}\hat{E}^{-1}$ and commutation relations (3), we derive the commutation relation for the velocity and boost momentum:

$$\left[\hat{v}_i, \hat{N}_j\right] = i\left(\delta_{ij} - \hat{P}_i\hat{P}_j\hat{E}^{-2}\right), \tag{27}$$

or $\left[\hat{v}_i, \hat{N}_i\right] = i\left(1-\hat{v}_i^2\right)$. This suggests that the uncertainty between the velocity and boost momentum is reduced for large relativistic velocities and their uncertainties (akin to the reduction of the angular-momentum uncertainty for large angle uncertainties [35,36]). Note also that for the 1D Klein-Gordon equation, $1-\hat{v}^2 = \mu^2\hat{E}^{-2}$, so that the velocity and boost momentum commute in the massless case $\mu = 0$ (all solutions propagate with the speed of light).

*3.6. Modes of the full Klein-Gordon equation*

Up to now, we considered the boost eigenmodes as functions of $(t,z)$, ignoring the transverse $(x,y)$ coordinates in the full Klein-Gordon equation (1). As it was mentioned in Section 2, there are two basic ways to satisfy the full Klein-Gordon equations using the Lorentz-boost eigenmodes. In the first approach, we multiply the boost modes (12) by unknown transverse functions $\chi(x,y)$:

$$\psi(t,\mathbf{r}) = \psi_n(t,z)\chi(x,y). \tag{28}$$

Substituting this into Eq. (1), and taking into account that $\psi_n(t,z)$ satisfy the 1D wave equation $\left(\partial_t^2 - \partial_z^2\right)\psi_n(z,t) = 0$, we find that $\chi(x,y)$ must satisfy the 2D Helmholtz-like equation:

$$\left(\nabla_\perp^2 - \mu^2\right)\chi(x,y) = 0. \tag{29}$$

According to Eqs. (3), the boost operator $\hat{N}_z$ commutes with the angular-momentum operator $\hat{L}_z$, and we can choose $\chi(x,y)$ to be vortex eigenmodes of $\hat{L}_z$. Using polar coordinates $(r_\perp,\varphi)$, we substitute $\chi_m(r_\perp,\varphi) = R_m(r_\perp)\exp(im\varphi)$ into Eq. (27), and obtain equation for the radial function:

$$R_m'' + \frac{1}{r_\perp}R_m' - \left(\mu^2 + \frac{m^2}{r_\perp^2}\right)R_m = 0. \tag{30}$$

Solutions of this equation are:

$$R_m \propto r^{\pm m} \text{ for } \mu = 0 \quad \text{and} \quad R_m \propto \left\{I_m(\mu r_\perp), K_m(\mu r_\perp)\right\} \text{ for } \mu > 0, \tag{31}$$

where $I_m$ and $K_m$ are the modified (hyperbolic) Bessel functions.

Thus, in contrast to cylindrical vortex beams (9) with bounded radial distributions (e.g., given by the regular Bessel functions $J_{|m|}(k_\perp r_\perp)$), the boost-vortex eigenmodes $\psi_{nm} = \psi_n(t,z)R_m(r_\perp)\exp(im\varphi)$ have radial distributions (28) *diverging either for $r_\perp \to 0$ or for $r_\perp \to \infty$*. (The only exclusion is the $m = 0$ mode with $\chi = $ const in the massless case $\mu = 0$.) Such divergent behavior is a consequence of the fact that we have constructed Lorentz-invariant signals propagating exactly with the speed of light for *massive* particles. The main part of a physical signal in the massive Klein-Gordon equation will actually propagate with a subluminal velocity and with significant distortions of its shape [21]. Nonetheless, the boost eigenmodes (12) only provide a basis, and one can construct subluminal physical signals by interfering boost modes with different signs of $n$. Moreover, some "unphysical" wave solutions, such as non-diffracting Bessel beams [30,37,38] or "accelerating" Airy beams [39–41] (both carrying infinite energy), have experimentally observable approximations in finite space-time domain.

It is worth remarking that Eq. (30) has well-defined localized solutions in the case of *tachyons* with imaginary mass, $\mu^2 < 0$. Then, the solutions are $R_m \propto J_{|m|}(i\mu r_\perp)$, and tachyons can be radially localized and propagate exactly with the speed of light.

The second approach to satisfy the Klein-Gordon equation is to ascribe the mass term to the longitudinal wave equation for $\psi_n$:

$$\left(\partial_t^2 - \partial_z^2 + \mu^2\right)\psi_n(t,z) = 0. \tag{32}$$

Obviously, the modes (12) do not satisfy it for $\mu \neq 0$, but these can be improved using the suitable multiplication (14): $\psi_n \to f(u)\psi_n$. Substituting Eqs. (12) and (14) into Eq. (32), we find that the Klein-Gordon equation (32) is satisfied by the function $f(u) = u^{|n|}\tilde{f}(u)$, where $\tilde{f}(u)$ is a solution of the Bessel equation

$$\partial_u^2 \tilde{f} + \frac{1}{u}\partial_u \tilde{f} + \left(\frac{n^2}{u^2} + \mu^2\right)\tilde{f} = 0. \tag{33}$$

Here we used the hyperbolic coordinates (13). This results in the Lorentz-boost modes $\psi_n(u,\phi) \to \tilde{f}(u)e^{in\phi}$, which were considered within the point-form quantum field theory approach [25–28]. These longitudinal wavefunctions can be multiplied by the transverse wavefunction satisfying the Helmholtz equation: $\nabla_\perp^2 \chi(x,y) = 0$, including $\chi(x,y) = \text{const}$. As we pointed out, the drawback of this approach is its nontrivial extension outside the forward light cone.

*3.7. Eigenmodes of the squared boost momentum*

The eigenmodes of $\hat{N}_z$ is related to Lorentz boosts along only one axis. Similarly to spherical eigenmodes of the squared angular momentum $\hat{L}^2$, playing a highly important role in physics [13], it is interesting find eigenmodes of the squared boost momentum $\hat{N}^2$. Note that $[\hat{L}^2, \hat{L}_z] = 0$, but $[\hat{N}^2, \hat{N}_z] \neq 0$, and one cannot construct eigenmodes of $\hat{N}^2$ and $\hat{N}_z$. However, $[\hat{N}^2, \hat{L}_z] = 0$ and $[\hat{N}^2, \hat{L}^2] = 0$, so it makes sense to seek eigenmodes of $\hat{N}^2$, $\hat{L}^2$ and $\hat{L}_z$ having the spherical-mode dependence on the spherical angles $(\vartheta, \varphi)$. Moreover, using the spherical coordinates $(r, \vartheta, \varphi)$, the squared boost-momentum operator (11) can be written as $\hat{N}^2 = -t^2 \nabla^2 - r^2 \partial_t^2 - r\partial_r - 3t\partial_t - 2tr\partial_t\partial_r$, where the Laplace operator can be presented as $\nabla^2 = r^{-1}\partial_r(r\partial_r) - r^{-2}\hat{L}^2$. Thus, we seek eigenmodes of $\hat{N}^2$ in the form:

$$\psi_{\nu\ell m} = F_\nu(t,r)Y_{\ell m}(\vartheta,\varphi), \quad Y_{\ell m}(\vartheta,\varphi) \propto P_\ell^{|m|}(\cos\vartheta)\exp(im\varphi). \tag{34}$$

Here, $Y_{\ell m}(\vartheta,\varphi)$ are the spherical eigenmodes of $\hat{L}^2$ and $\hat{L}_z$, with the eigenvalues $\ell(\ell+1)$ and $m$, respectively [13], $P_\ell^{|m|}$ are associated Legendre polynomials, while $F_\nu(t,r)$ is an eigenmode of $\hat{N}^2$ with the eigenvalue $\nu$. Substituting the wavefunction (34) into the eigenmodes equation $\hat{N}^2 \psi_{\nu\ell m} = \nu \psi_{\nu\ell m}$ and the Klein-Gordon equation (1), we find that the function $F_\nu(t,r)$ must satisfy two differential equations:

$$\left\{-t^2\left[r^{-1}\partial_r(r\partial_r) - \ell(\ell+1)r^{-2}\right] - r^2\partial_t^2 - r\partial_r - 3t\partial_t - 2tr\partial_t\partial_r - \nu\right\}F_\nu = 0,$$

$$\left\{r^{-1}\partial_r(r\partial_r) - \partial_t^2 - \ell(\ell+1)r^{-2} - \mu^2\right\}F_\nu = 0. \tag{35}$$

The variables are not separated here, and we could not find analytical solutions of these equations. Finding these eigenmodes, if they exist, is a problem for future study.

## 4. Discussion

To conclude, we have examined eigenmodes of the boost-momentum operator, i.e., a generator of Lorentz boosts. Our modes are different from the boost eigenmodes considered in the point-form quantum field theory [25–28]. The modes considered in this paper are simple functions of $t-z$ or $t+z$, satisfying the 1D massless wave equation, while the previously-considered modes are more complicated functions, initially defined only inside the forward light cone, and satisfying the 1D Klein-Gordon equation. Therefore, we argue that our eigenmodes, looking like step signals propagating with the speed of light, could be more suitable for the causal signal-propagation problems. The local frequency (wave vector) and particle density diverge near the signal front and decay away from it. This makes the boost modes scale-invariant and invariant under Lorentz boosts. Constructing a Gaussian-weighted superposition of boost modes with different eigenvalues removes the wavefunction discontinuity and the particle-density divergence, and allows one to trace transition from step-like signals to Gaussian-like wave packets (smooth but non-analytical at the signal front).

The boost modes should be compared with eigenmodes of other generators of the Poincaré group: plane waves (the energy-momentum eigenmodes) and cylindrical or spherical vortex modes (the angular momentum eigenmodes). Each of these sets is suitable for certain kinds of problems involving the corresponding translational or rotational symmetries. The boost eigenmodes are intimately related to the Lorentz-transformation symmetry, which is usually less important for physical applications. A distinctive feature of the boost modes is that they describe propagating relativistic *signals*. For example, the wavefunction can be chosen to vanish for $z > t$ and to be non-zero for $z < t$. In contrast, even localized (square-integrable) superpositions of plane waves or vortex modes spread over all spacetime, and it this feature that results in the counterintuitive "superluminal" propagation of Gaussian wave packets [22–24]. Therefore, the boost eigenmodes can play an important role in problems involving *causality* and signal propagation [19–21].

The Lorentz-boost modes exhibit exotic wave forms, which look similar in any reference frame, i.e., a moving observer will receive exactly the same signal independently of its motion. (For plane waves, the observer/source motion results in the blue or red frequency shift.) The uncertainty (commutation) relations between the boost momentum and momentum/velocity reveal important features of the boost eigenmodes. Pure modes have well-defined boost momenta and uncertain momenta/velocities, because all frequencies or wave vectors are present in these modes. Considering superpositions of multiple boost modes with large uncertainty in their eigenvalues, we arrive at Gaussian-like wave packets with well-pronounced central frequency and momentum. Such wave packets will look different in different reference frames due to the blue/red shifts of the central frequency.

Since the simplest boost eigenmodes (12), mostly considered in this paper, are solutions of 1D massless wave equation, considering them in the context of 3D Klein-Gordon equation meets some difficulties. Namely, the corresponding wave functions of the transverse (cylindrical) coordinates $(r_\perp, \varphi)$ become divergent either for $r_\perp = 0$ or for $r_\perp = \infty$. This is because massive particles cannot form perfect signals propagating with the speed of light. Even massless waves in 3D can propagate with the speed of light along the $z$-axis only if these are delocalized in the transverse plane. In contrast, the modified boost eigenmodes (14), (32), (33) considered in [25–28] are non-divergent, but propagate with significant shape deformations. In any case, for causal signal propagation in the massive Klein-Gordon equation, there is always a *Sommerfeld precursor*, which propagates exactly with the speed of light (although the main part of the signal propagates with a subluminal group velocity) [19–21]. This corresponds to the fact that the signal propagation is always limited by the speed of light, with the precursor travelling exactly at this speed, even when the group velocity is super- or sub-luminal. In agreement with this, the boost eigenmodes exhibit the same speed-of-light signal-front propagation independently of the mass, including the imaginary-mass case of *tachyons* (which are characterized by a superluminal group velocity).

We finally note that the boost modes carry essential *wavefunction singularities*, which can be important in the general context of wave physics. Indeed, smooth wave fields generically possess only *phase singularities* (vortices), i.e., points of zero amplitude and indeterminate phase [16–18]. These singularities play crucial roles in various problems, and they underpin the angular momentum eigenmodes [14,15,30]. In contrast, the signal-propagation problems require the wavefunction to vanish in a finite spacetime domain, which brings about stronger singularities in the amplitude and phase, such as step functions with divergent phase gradients. It is possible that such singularities always propagate with the speed of light, independently of the mass, because a divergent phase gradient implies a divergent local momentum, whereas the relativistic dispersion $E = \sqrt{P^2 + \mu^2}$ corresponds to the luminal propagation for $P \to \infty$. In fact, the Sommerfeld signal precursors, travelling with the speed of light, exhibit wave forms resembling the boost-mode singularities, with the local frequency of oscillations increasing near the front [19–21].

## Acknowledgements

I am grateful to M. V. Berry, Y. P. Bliokh, and E. A. Ostrovskaya for helpful discussions and important corrections. I am also indebted to an anonymous referee for drawing my attention to the relevant quantum-field-theory works [25–28]. This work was supported by the Australian Research Council.

# References

1. E. P. Wigner, *Symmetries and Reflections* (Indiana University Press, 1967).
2. L. D. Landau and E. M. Lifshitz, *Mechanics*, 3rd ed. (Butterworth-Heinemann, 1976).
3. D. E. Soper, *Classical Field Theory* (John Wiley & Sons, 1976).
4. H. M. Schwartz, *Introduction to Special Relativity* (McGraw-Hill, 1968)
5. I. Bialynicki-Birula and Z. Bialynicki-Birula, *Quantum Electrodynamics* (Pergamon Press, 1975).
6. L. D. Landau and E. M. Lifshitz, *Classical Theory of Fields*, 4th ed. (Butterworth-Heinemann, 1994).
7. S. M. Barnett, On the six components of optical angular momentum, *J. Opt.* **13**, 064010 (2011).
8. K. Y. Bliokh and F. Nori, Relativistic Hall effect, *Phys. Rev. Lett.* **108**, 120403 (2012).
9. K. Y. Bliokh, A. Y. Bekshaev, and F. Nori, Dual electromagnetism: helicity, spin, momentum, and angular momentum, *New J. Phys.* **15**, 033026 (2013); Corrigendum: *New J. Phys.* **18**, 089503 (2016).
10. R. P. Cameron and S. M. Barnett, Electric-magnetic symmetry and Noether's theorem, *New J. Phys.* **14**, 123019 (2012).
11. I. Bialynicki-Birula and Z. Bialynicki-Birula, Canonical separation of angular momentum of light into its orbital and spin parts, *J. Opt.* **13**, 064014 (2011).
12. D. A. Smirnova, V. M. Travin, K. Y. Bliokh, and F. Nori, Relativistic spin-orbit interactions of photons and electrons, *Phys. Rev. A* **97**, 043840 (2018).
13. L. D. Landau and E. M. Lifshitz, *Quantum Mechanics*, 3rd ed. (Butterworth-Heinemann, 1977).
14. L. Allen, S. M. Barnett, and M. J. Padgett, *Optical Angular Momentum* (Taylor and Francis, 2003).
15. D. L. Andrews and M. Babiker, *The Angular Momentum of Light* (Cambridge University Press, 2012).
16. J. F. Nye and M. V. Berry, Dislocations in wave trains, *Proc. R. Soc. Lond. A* **336**, 165 (1974).
17. M. S. Soskin and M. V. Vasnetsov, Singular optics, *Prog. Opt.* **42**, 219 (2001).
18. M. R. Dennis, K. O'Holleran, and M. J. Padgett, Singular optics: optical vortices and polarization singularities, *Prog. Opt.* **53**, 293 (2009).
19. A. Sommerfeld, *Optics* (Academic Press, 1950).
20. L. Brillouin, *Wave Propagation and Group Velocity* (Academic Press, 1960).
21. M. V. Berry, Causal wave propagation for relativistic massive particles: physical asymptotics in action, *Eur. J. Phys.* **33**, 279 (2012).
22. R. Y. Chiao and A. M. Steinberg, Tunneling times and superluminality, *Prog. Opt.* **37**, 345 (1997).
23. H. G. Winful, Tunneling time, the Hartman effect, and superluminality: A proposed resolution of an old paradox, *Phys. Rep.* **436**, 1 (2006).
24. M. Asano, K. Y. Bliokh, Y. P. Bliokh, A. G. Kofman, R. Ikuta, *et al.*, Anomalous time delays and quantum weak measurements in optical micro-resonators, *Nature Commun.* **7**, 13488 (2016).
25. C. M. Sommerfield, Quantization on spacetime hyperboloids, *Ann. Phys.* **84**, 285 (1974).
26. D. Gromes, H. J. Rothe, and B. Stech, Field quantization on the surface $X^2 = \text{constant}$, *Nucl. Phys. B* **75**, 313 (1974).
27. A. diSessa, Quantization on hyperboloids and full space-time field expansion, *J. Math. Phys.* **15**, 1892 (1974).
28. E. P. Biernat, W. H. Klink, W. Schweiger, and S. Zelzer, Point-form quantum field theory, *Ann. Phys.* **323**, 1361 (2008).

29. V. B. Berestetskii, E. M. Lifshitz, and L.P. Pitaevskii, *Quantum Electrodynamics* (Butterworth-Heinemann, 1982).
30. K. Y. Bliokh, I. P. Ivanov, G. Guzzinati, L. Clark, R. Van Boxem, A. Béché, R. Juchtmans, M. A. Alonso, P. Schattschneider, F. Nori, and J. Verbeeck, Theory and applications of free-electron vortex states, *Phys. Rep.* **690**, 1 (2017).
31. B. Thaller, *Visual Quantum Mechanics* (Springer, 2000).
32. M. V. Berry, Five momenta, *Eur. J. Phys.* **34**, 1337 (2013).
33. A. Messiah, *Quantum Mechanics* (North-Holland, 1962).
34. B. Thaller, *The Dirac Equation* (Springer, 1992).
35. S. M. Barnett and D. T. Pegg, Quantum theory of rotation angles, *Phys. Rev. A* **41**, 3427 (1990).
36. S. Franke-Arnold, S. M. Barnett, E. Yao, J. Leach, J. Courtial, and M. Padgett, Uncertainty principle for angular position and angular momentum, *New J. Phys.* **6**, 103 (2004).
37. J. Durnin, J. J. Miceli, and J. H. Eberly, Diffraction-free beams, *Phys. Rev. Lett.* **58**, 1499 (1987).
38. V. Grillo, E. Karimi, G. C. Gazzadi, S. Frabboni, M. R. Dennis, and R. W. Boyd, Generation of nondiffracting electron Bessel beams, *Phys. Rev. X* **4**, 011013 (2014).
39. M. V. Berry and N. L. Balazs, Nonspreading wave packets, *Am. J. Phys.* **47**, 264 (1979).
40. G. A. Siviloglou, J. Broky, A. Dogariu, and D. N. Christodoulides, Observation of accelerating Airy beams, *Phys. Rev. Lett.* **99**, 213901 (2007).
41. N. Voloch-Bloch, Y. Lereah, Y. Lilach, A. Gover, and A. Arie, Generation of electron Airy beams, *Nature* **494**, 331 (2013).